\documentclass[12]{article}

\usepackage{amsmath,amssymb,amsthm}
\usepackage{hyperref}
\usepackage{natbib}
\usepackage{mathtools}
\usepackage[title]{appendix}
\usepackage{fullpage}
\usepackage{setspace,rotating}
\usepackage{threeparttable,booktabs}

\newcommand{\E}{\mathrm{E}}
\newcommand{\Var}{\mathrm{Var}}

\hypersetup{
	colorlinks   = true, 
	urlcolor     = blue, 
	linkcolor    = blue, 
	citecolor   = blue 
}

\begin{document}
	
	\title{The Yule-Frisch-Waugh-Lovell Theorem}
	
	\author{Deepankar Basu\thanks{Department of Economics, University of Massachusetts Amherst. Email: \texttt{dbasu@econs.umass.edu}.}}
	
	\date{\today}
	
	\maketitle
	
	\begin{abstract}
		This paper traces the historical and analytical development of what is known in the econometrics literature as the Frisch-Waugh-Lovell theorem. This theorem demonstrates that the coefficients on any subset of covariates in a multiple regression is equal to the coefficients in a regression of the residualized outcome variable on the residualized subset of covariates, where residualization uses the complement of the subset of covariates of interest. In this paper, I suggest that the theorem should be renamed as the Yule-Frisch-Waugh-Lovell (YFWL) theorem to recognize the pioneering contribution of the statistician G. Udny Yule in its development. Second, I highlight recent work by the statistician, P. Ding, which has extended the YFWL theorem to a comparison of estimated covariance matrices of coefficients from multiple and partial, i.e. residualized regressions. Third, I show that, in cases where Ding's results do not apply, one can still resort to a computational method to conduct statistical inference about coefficients in multiple regressions using information from partial regressions.
		
	\end{abstract}
	
	\doublespacing
	
	\section{Introduction}
	
	The Frisch-Waugh-Lovell theorem is a remarkable result about linear regressions estimated with the method of least squares. The theorem shows that coefficients of variables in a multiple regression are exactly equal to corresponding coefficients in partial regressions that use residualized versions of the dependent and independent variables. While it has been primarily used in econometrics \citep{krishnakumar-2006,tielens-vanhove-2017}, it has found wider applications in a variety of disciplines, including statistics \citep{arendacka-puntanen-2015,gross-moeller-2023}, electrical engineering \citep{monsurro-trifiletti-2017}, computer science \citep{ahrens-etal-2021}, and genetics \& molecular biology \citep{reeves-etal-2012}, to name just a few. It is now included in many graduate-level textbooks in econometrics, including \citet[p.~19--24]{davidson-mackinnon}, \citet[p.~73--74]{hayashi}, \citet[pp.~35--36]{angrist-pischke-2009}, and \citep[p.~33]{greene}. 
	
	In the econometrics literature, the theorem is understood to have originated in a $ 1933 $ paper by Ragnar Frisch and Frederick V. Waugh in the first volume of  \textit{Econometrica} \citep{frisch-waugh-1933}, which was later generalized by Michael C. Lovell \citep{lovell-1963}; hence the name Frisch-Waugh-Lovell theorem. The first contribution of this paper is to show that the result was proved more than two and a half decade ago by the statistician G. Udny Yule in a $ 1907 $ paper \citep{udny-1907}. This seems to be well known in statistics \citep[p.~60]{agresti-2015} and should be recognized in econometrics as well. In fact, to recognize Yule's pioneering contribution to the development of this important result, we should refer to this important result as the Yule-Frisch-Waugh-Lovell (YFWL) theorem, rather than the currently used Frisch-Waugh-Lovell (FWL) theorem.
	
	The second contribution of this paper is to trace out the analytical \textit{development} of the theorem through the decades. In \citet{udny-1907}, who proved the result using basic algebra, the partial regressions are always bivariate regressions of residual vectors. In \citet{frisch-waugh-1933}, where the proof relied on some basic properties of determinant and Cramer's rule for finding solutions of linear systems of equations, the partial regressions are themselves multiple regressions but the residuals are computed with bivariate regressions. Thus, \citet{udny-1907} allows multiple variables in the conditioning set, but conceives of the partial regressions as bivariate regressions. On the other hand, \citet{frisch-waugh-1933} allows only one variable (the linear time trend) in the conditioning set, but allows the partial regressions to be multiple regressions. \citet{lovell-1963} generalizes in both directions, allowing multiple variables in the conditioning set and allowing the partial regressions to themselves be multiple regressions.\footnote{\citet[p.~1000]{lovell-1963} points out that \citet[p.~304]{tintner-1957} had extended the Frisch-Waugh result to the case of polynomial time trends.} In terms of methodology, \citet{lovell-1963} introduces the use of projection matrices from linear algebra, making the proof compact and elegant. Current presentations of the theorem follow Lovell's exposition \citep{davidson-mackinnon,greene}. 
	
	One use of the YFWL theorem is that if researchers are interested in only a subset of coefficients in a multiple regression equation, they need not estimate the full multiple regression. They can estimate the relevant partial regressions to recover estimates of the desired coefficients. In contexts of empirical analyses with large numbers of variables and observations, this can significantly reduce computational time and resources. Will researchers also be able to conduct inference about the subset of coefficients in a multiple regression using the standard errors (or the covariance matrix) estimated from the relevant partial regression? \citet{udny-1907} and \citet{frisch-waugh-1933} did not pose, and therefore did not address, this question about standard errors (or the covariance matrix). 
	
	\citet{lovell-1963} addressed the issue of the covariance matrices very briefly by pointing out that \textit{estimated} standard errors from the multiple and partial regressions are equal up to a degree of freedom adjustment when the true error is homoskedastic. In a recent contribution, \citet{ding-2021} has analyzed the question of estimated standard error equivalence between multiple and partial regressions in greater detail. \citet{ding-2021} has demonstrated that estimates of the covariance matrices from the multiple and partial regressions are equal up to a degree of freedom adjustment in the case of homoskedastic errors and, quite surprisingly, are exactly equal for some variants of heteroskedasticity consistent covariance matrices (HCCM), heteroskedasticity and autocorrelation consistent (HAC) covariance matrices, and for some variants of clustered robust covariance matrices. 
	
	The analysis in \citet{ding-2021} highlights a general principle: if the estimate of error covariance matrix does not depend on the matrix of covariates beyond its dimensions, then estimated covariance matrices from multiple and partial regressions are equal, either exactly or with a degree of freedom correction. This principle can be leveraged to offer a computational solution for cases where \citet{ding-2021}'s results do not hold, i.e. estimates and other information from partial regressions are sufficient to compute correct covariance matrices of coefficient vectors in the multiple regression. If only a subset of coefficients is of interest then estimation and proper statistical inference can be conducted by working with the partial regression alone. In cases where the number of regressors is large, this might reduce computational burden.
	
	The rest of the paper is organized as follows: in section~\ref{sec:setup}, I introduce the set-up and pose the main question; in section~\ref{sec:yule}, I present Yule's results; in section~\ref{sec:frischwaugh}, I discuss Frisch and Waugh's contribution; in section~\ref{sec:lovell}, I discuss how Lovell extended the previous analysis; in section~\ref{sec:ding}, I present Ding's extension of the previous literature; I conclude in section~\ref{sec:conclusion} with two observations about the analytical and historical development of the YFWL theorem. Proofs are collected together in the appendix.  
	
	\section{The set-up and the question}\label{sec:setup}
	
	Consider a linear regression of an outcome variable, $Y$, on a set of $k$ covariates, 
	\begin{equation}\label{model-full}
		Y = W \beta + \varepsilon,
	\end{equation}
	where $Y$ is a $N \times 1$ vector of the outcome variable, $W$ is a $N \times k$ matrix of covariates (with the first column being a column of $1$s to capture the constant term in the regression function), $\beta$ is a $k \times 1$ of parameters, and $\varepsilon$ is a $N \times 1$ vector of the stochastic error term. 
	
	Suppose on the basis of a specific question under investigation, it is possible for the researcher to partition the set of regressors into two groups, $W_1$ and $W_2$, where the former is a $N \times k_1$ matrix (with the first column being a column of $1$s)  and the latter is a $N \times k_2$ matrix, with $k=k_1+k_2$. The model in (\ref{model-full}) can then be written as 
	\begin{equation}
	Y = W_1 \beta_1 + W_2 \beta_2 + \varepsilon,
	\end{equation}
	where $\beta_1$ and $\beta_2$ are $k_1 \times 1$ and $k_2 \times 1$ vectors of parameters. 
	
	Suppose the researcher is only interested in the first group of regressors ($W_1$), i.e. she is only interested in estimating and conducting inference about $\beta_1$. But, of course, she does not want to completely throw away the information in $W_2$ because the variables in this subset do impact $Y$ and are likely to also be correlated to the variables in $W_1$. Hence, she wants to condition her analysis on $W_2$, but is not interested in their partial effects on the outcome.
	
	Given her interest and the set-up, can the researcher avoid estimating the full model in (\ref{model-full})? Can she work with a smaller, let us say partial, model that allows her to consistently estimate and conduct proper statistical inference on $\beta_1$? The YFWL theorem provides an answer in the affirmative.
	
	\section{Pioneering contribution of Yule}\label{sec:yule}
	
	\subsection{Novel notation}
	
	\citet{udny-1907} considers the relationship among $n$ random variables $X_1$, $ X_2$, $\ldots$, and $X_n$ using a sample which has $N$ observations on each of the variables. In particular, suppose this relationship is captured by a linear regression of the first variable, $X_1$, on the other variables, $X_2, \ldots, X_n$, and a constant. This linear regression is what Yule called a \textit{multiple regression}. With reference to (\ref{model-full}), therefore \citet{udny-1907} uses: $k=n$, $X_1=Y$, $X_2$ the second column of $W$, $X_3$ the third column of $W$, and so on.
	
	Subtracting the sample mean from all variables, the multiple regression equation can be equivalently expressed in `deviation from mean' form in which the constant drops out. Let $x_1, x_2, \ldots, x_n$ denote the same random variables, but now expressed as deviations from their respective sample means. Introducing a novel notation to express the coefficients and the residual, \citet{udny-1907} writes the sample analogue of the multiple regression function as follows:
	\begin{equation}\label{lrg-1}
		x_1 = \underbrace{b_{12.34 \ldots n} x_2 + b_{13.24 \ldots n} x_3 + \cdots + b_{1n.234 \ldots n-1} x_n}_{\text{predicted value}} + \underbrace{x_{1.234 \ldots n}}_{\text{residual}}.
	\end{equation}	 
	
	In (\ref{lrg-1}), the subscripts of the coefficient on each regressor, as also the residual, is divided into the primary and secondary subscripts. The primary subscripts come before the period; the secondary subscripts come after the period. For the coefficients of regressors, the primary subscripts identify the dependent variable and that particular regressor, in that order; and the secondary subscripts identify the other regressors in the model. For the residual in (\ref{lrg-1}), $x_{1.234 \ldots n}$, the primary subscript identifies the dependent variable and the secondary subscripts identify all the regressors.
	
	It is useful to note three features of the subscripts. First, the order of the primary subscripts is important but the order among the secondary ones are not because the order in which the regressors are arranged in immaterial for estimation and inference. For instance, $b_{12.34 \ldots n}$ denotes the coefficient on $x_2$ for a regression of $x_1$ on $x_2, x_3, \ldots, x_n$. Similarly, $b_{1j.234 \ldots n}$ denotes the coefficient on $x_j$ for a regression of $x_1$ on $x_2, x_3, \ldots, x_n$, and note that the secondary subscripts excludes $ j $. Second, elements of the subscripts cannot be repeated because it is not meaningful to include the dependent variable as a regressor or repeat a regressor itself. Third, since the coefficients of the multiple regression function are related to partial correlations, the notation was also used for denoting relevant partial correlations too.\footnote{See, for instance, equation (2) in \citet{udny-1907}.}

	
	\subsection{The theorem}
	
	Now consider, with reference to the multiple regression (\ref{lrg-1}), what Yule called \textit{partial regressions}. These refer to regressions with partialled out variables. For instance, with reference to (\ref{lrg-1}), the partial regression of $x_1$ on $x_2$ would be constructed as follows: (a) first run a regression of $x_1$ on all the variables other than $x_2$, i.e. $x_3, x_4, \ldots, x_n$, and collect the residuals; (b) run a regression of $ x_2 $ on all the other regressors, i.e. $x_3, x_4, \ldots, x_n$, and collect the residuals; (c) run a regression of the residuals from the first step (partialled out outcome variable) on the residuals from the second step (partialled out first regressor). 
	
	\citet[p.~184]{udny-1907} showed that, if parameters of the multiple regression and the partial regression(s) were estimated by the method of least squares then the corresponding coefficients would be numerically the same. For instance, considering the first regressor, $x_2$, in (\ref{lrg-1}), he demonstrated that
	\begin{equation}\label{yule-1}
		\frac{\sum \left( x_{1.34 \ldots n} \times x_{2.34 \ldots n}\right) }{\sum \left( x_{2.34 \ldots n}\right)^2 } = b_{12.34 \ldots n}  
	\end{equation}
	where the summation runs over all observations in the sample.\footnote{For a proof see appendix~\ref{app:yule}.}
	
	To see why (\ref{yule-1}) gives the claimed result, recall that $x_{1.34 \ldots n}$ is the residual from a regression of $ x_1 $ on $ x_3, \ldots, x_n $ and $x_{2.34 \ldots n}$ is the residual from the regression of $ x_2 $ on $ x_3, \ldots, x_n $. By construction, both have zero mean in the sample. Recall, further, that in a regression of a zero-mean random $y$ on another zero-mean random variable $z$, the coefficient on the latter is $\sum y z /\sum z^2$. Hence, the left hand side of (\ref{yule-1}) is the coefficient in a regression of $x_{1.34 \ldots n}$ on $x_{2.34 \ldots n}$. The right hand side is, of course, the coefficient on $x_2$ in (\ref{lrg-1}). Hence, the result.
	
	The above argument, of course, can be applied to any of the regressors in (\ref{lrg-1}) and thus \citet{udny-1907} proved the following general result: the coefficient on any regressor in (\ref{lrg-1}) is the same as the coefficient in a bivariate regression of the residualized dependent variable (the vector of residual from a regression of the dependent variable on the other regressors) and the residualized regressor (the vector of residuals from a regression of the regressor of interest on the other regressors). 
	
	With reference to the question posed in section~\ref{sec:setup}, \citet{udny-1907}, therefore, showed that a researcher could work with partial regressions if she were only interested in a subset of the coefficients in the multiple regression. She would arrive at the same estimate of the parameters by estimating partial regressions as she would if she had estimated the full model. In particular, when $W_1$ contained only one variable (other than the column of $1$s), \citet{udny-1907} provided the following answer: (a) run a regression of $y$ (demeaned $Y$) on the variables in $w_2$ (demeaned variables in $W_2$), and collect the residuals; (b) run a regression of $w_1$ (demeaned variable in $W_1$, i.e. excluding the constant) on the variables in $w_2$, and collect the residuals; (c) regress the first set of residuals on the second to get the desired coefficient.
	
	There are two things to note about Yule's answer. First, he did not provide an answer to the question posed in section~\ref{sec:setup} when $W_1$ contained more than one random variable (excluding the constant). Of course, partial regressions could be estimated for each independent variable, but they had to be separately estimated as bivariate regressions with \textit{all} the other variables used for the partial regressions. Second, \citet{udny-1907} did not investigate the relationship among variances of the parameters estimated from multiple and partial regressions. Is the \textit{estimated} standard error of a coefficient also identical from the full and partial regressions? Yule did not pose, and hence did not provide any answers to, this question. 
	
	\subsection{Application to a regression with a linear time trend}
	
	We can immediately apply the result from Yule's theorem given above to a regression with a linear time trend. With (\ref{lrg-1}), let $x_n$ denote the demeaned linear time trend variable. Applying (\ref{yule-1}), we will have the following result: the coefficient on any of the regressors, $x_2, x_3, x_{n-1}$ in (\ref{lrg-1}) is the same as the coefficient in a bivariate regression of residualized $x_1$ on the relevant regressor. This is of course the exact same result that was presented $26$ later in \citet{frisch-waugh-1933}. 
	
	\section{Frisch and Waugh's result}\label{sec:frischwaugh}
	
	\subsection{Notation and result}
	\citet[p.~389]{frisch-waugh-1933} study the relationship among $n+1$ variables, one of which is a linear time trend. In reference to (\ref{model-full}), therefore \citet{frisch-waugh-1933} use: $k=n$, $X_0=Y$, $X_1$ the second column of $W$, $X_2$ the third column of $W$, and so on. They use the same convention of considering variables in the form of deviations from their respective sample means.\footnote{``Consider now the $ n $ variables $x_0, \ldots, x_{n-1}$ and let time be an $ (n+1)$th variable $x_n$. Let all the variables be measured from their means so that $\sum x_i=0 \quad (i = 0, \ldots, n)$ where $\sum$ denotes a summation over all the observations.'' \citep[p.~394]{frisch-waugh-1933}.} Using the \textit{exact same notation} as used by \citet{udny-1907} and postulating an \textit{a priori} true linear relationship among these variables, they write,
	\begin{equation}\label{fw-1}
		x_0 = \beta_{01.34 \ldots n} x_1 + \beta_{02.24 \ldots n} x_2 + \cdots + \beta_{0n.234 \ldots n-1} x_n,
	\end{equation}
	and consider estimating the parameters of (\ref{fw-1}) in two ways. First, they consider the multiple regression of $x_0$ on  $x_1, \ldots, x_n$:
	\begin{equation}\label{fw-2}
		x_0 = b_{01.34 \ldots n} x_1 + b_{02.24 \ldots n} x_2 + \cdots + b_{0n.234 \ldots n-1} x_n.
	\end{equation}
	Second, they consider the multiple regression of $x'_0$ on  $x'_1, \ldots, x'_{n-1}$ (note that $x_n$ has been excluded from the set of regressors),
	\begin{equation}\label{fw-3}
		x'_0 = b'_{01.34 \ldots n} x'_1 + b'_{02.24 \ldots n} x'_2 + \cdots + b'_{0,n-1.234 \ldots n-1} x'_{n-1},
	\end{equation}
	where the primed variables are the corresponding time-demeaned original variables, i.e., for $j = 0, 1, \ldots, n-1$, $x'_j$ is the residual in the regression of $x_j$ on $x_n$.\footnote{Note that (\ref{fw-1}), (\ref{fw-2}) and (\ref{fw-3}) are just the predicted regressions functions. These equations exclude the residuals.} Using the basic rules of determinants and Cramer's rule for solving linear equation systems, \citet[p.~394--396]{frisch-waugh-1933} demonstrated that the coefficients denoted by $b$ in (\ref{fw-2}) are numerically equal to the corresponding coefficients denoted by $b'$ in (\ref{fw-3}).\footnote{For a proof see appendix~\ref{app:frisch}.} 
	
	With regard to the question posed in section~\ref{sec:setup}, Frisch and Waugh provide the same answer as Yule: coefficients are the same whether they are estimated from the multiple or from partial regressions. There are both similarities and differences between \citet{udny-1907}  and \citet{frisch-waugh-1933}. First, whereas in \citet{udny-1907}, only one variable could be included in $W_1$ (the subset of covariates that was of interest to the researcher), in \citet{frisch-waugh-1933} only one random variable could be included in  $W_2$ (the subset of covariates that was \textit{not} of interest to the researcher). Second, much like \citet{udny-1907} before them, \citet{frisch-waugh-1933} did not investigate the relationship among estimated variances of the parameters of multiple and partial regressions. The question that is relevant for statistical inference, i.e. standard errors, had still not been posed.
	
%
	

	\section{Lovell extends the analysis}\label{sec:lovell}
	
	\citet{lovell-1963} extended the reach of the theorem significantly and addressed both questions that had been left unanswered by \citet{udny-1907} and \citet{frisch-waugh-1933}. On the one hand, \citet{lovell-1963} partitioned $W$ into two subsets without any restrictions on the number of variables in each; on the other, he laid the groundwork for thinking about the estimated covariance matrices of the coefficient vectors.
	
	\subsection{The results}
	To understand the argument in \citet{lovell-1963}, we can directly work with (\ref{model-full}). Consider the sample analogue of (\ref{model-full}), 
	\begin{equation}
		Y = W_1 b_1 + W_2 b_2 + u,
	\end{equation}
	where $b_1$ and $b_2$ are OLS estimators of $\beta_1$ and $\beta_2$, respectively, and $u$ is the residual (sample analogue of the error term, $\varepsilon$). To facilitate the algebra, we will rewrite the above as
	\begin{equation}\label{model-full-sample}
		Y = W_2 b_2 + W_1 b_1 + u,	
	\end{equation}
	where I have switched the position of the regressors of interest, $W_1$, and of those that are used merely for conditioning, $W_2$.
	
	Let $Y^*$ be the $N \times 1$ vector of residuals from a regression of $Y$ on $W_2$, and $W_1^*$ be the $N \times k_1$ matrix formed by column-wise stacking of the vectors of residuals obtained from regressing each column of $W_1$ on $W_2$, i.e. 
	\begin{equation}\label{defw1star}
		Y^* = \left[ I - W_2 \left( W'_2 W_2 \right) ^{-1} W'_2\right] Y \text{ and } W^*_1 = \left[ I - W_2 \left( W'_2 W_2 \right) ^{-1} W'_2\right] W_1.
	\end{equation}
	
	The matrices $P_{W_2}=W_2 \left( W'_2 W_2 \right) ^{-1} W'_2$ and $M_{W_2}$ play important roles in the whole analysis. They are both symmetric and idempotent. They are referred to, respectively, as the `hat matrix' and the `residual maker matrix'. This terminology comes from the fact that $P_{W_2}$ projects any vector in $\mathbb{R}^N$ onto the column space of $W_2$, and $M_{W_2}$ projects onto the orthogonal complement of the column space of $W_2$. Thus, $P_{W_2}$ generates the vector of predicted values and $M_{W_2}$ generates the vector of least squares residuals for a regression of any vector in $\mathbb{R}^N$ on $W_2$. This result holds for any regression, so that, for instance, the vector of residuals in (\ref{model-full-sample}) is given by $u = (I - P_W) Y$. 
	
	Now consider the regression of $Y^*$ on $W_1^*$:
	\begin{equation}\label{model-partial}
		Y^* =  W_1^* \tilde{b}_1 + \tilde{u},
	\end{equation}
	and note that the vector of residuals in (\ref{model-partial}) is given by $\tilde{u} = (I - P_{W^*_1}) Y^*$.
	
	\citet{lovell-1963} demonstrated two important results: (a) the coefficient vectors are numerically the same whether they are estimated from the multiple or the partial regressions, i.e. $b_1$ in (\ref{model-full-sample}) is exactly equal to  $\tilde{b}_1$ in (\ref{model-partial}), and (b) the vector of residuals from the multiple and partial regressions are numerically the same, i.e. $u$ in (\ref{model-full-sample}) is equal to $\tilde{u}$ in (\ref{model-partial}). The first result completed the YFWL so far as the estimate of the coefficient is concerned because the partitioning of the set of regressors was completely general; the second result laid the groundwork for comparing estimated variances of the coefficient vectors from multiple and partial regressions.\footnote{For a proof of Lovell's results, see appendix~\ref{app:lovell}. It is straightforward to extend the YFWL to generalized least squares \citep[p.~1004]{lovell-1963}. Other scholars have worked on variations of Lovell's results; see, for instance, \citet{fiebig-bartels-1996, krishnakumar-2006}.}
	
	\section{Ding considers standard errors}\label{sec:ding}
	
%
	
	The analysis so far has focused on comparing the coefficient vectors from multiple and partial regressions, i.e. $b_1$ from (\ref{model-full-sample}) and $\tilde{b}_1$ from (\ref{model-partial}). The YFWL theorem has established that they are numerically the same. This is a very useful result but still does not answer the question that would be relevant if a researcher were interested in conducting statistical inference about $b_1$. To conduct inference about $b_1$ using results of estimating $\tilde{b}_1$, we also need to be able to compare their estimated covariance matrices. 
	
	With reference to the multiple regression (\ref{model-full-sample}), the estimated variance of the coefficient vector is
	\begin{equation}
		\widehat{\Var} \begin{pmatrix}
			b_2 \\
			b_1
		\end{pmatrix} =
		\begin{pmatrix}
			\widehat{\Var} \left( b_2\right)  & \widehat{\E} \left( b_2, b'_1 \right) \\
			\widehat{\E} \left( b_1, b'_2 \right) & \widehat{\Var} \left( b_1\right)  
		\end{pmatrix} =
	\left(W' W\right)^{-1} W' \widehat{\Omega}_m W \left(W' W\right)^{-1},
	\end{equation}
	where $W = \left[ W_2 : W_1\right]$ and $\widehat{\Omega}_m$ (subscript `m' for identifying the multiple regression) is an estimate of the variance of the error term in the multiple regression model (\ref{model-full-sample}). Hence, the estimated variance of $b_1$ is  
	\begin{equation}\label{estvar-model-full}
		\widehat{\Var} \left( b_1\right) = \text{ (2,2) block of }\left(W' W\right)^{-1} W' \widehat{\Omega}_m W \left(W' W\right)^{-1}.
	\end{equation}
	Similarly, with reference to the partial regression (\ref{model-partial}), the estimated variance of $\tilde{b}_1$ is given by 
	\begin{equation}\label{estvar-model-partial}
	\widehat{\Var} \left( \tilde{b}_1\right) =	\left( {W^*_1}^{'} W^*_1\right)^{-1} {W^*_1}^{'} \widehat{\Omega}_p W^*_1 \left( {W^*_1}^{'} W^*_1\right)^{-1} . 
	\end{equation}
	where $W^*_1$ is defined (\ref{defw1star}), and $\widehat{\Omega}_p$ (subscript `p' for identifying the partial regression) is an estimate of the variance of the error term in the partial regression model (\ref{model-partial}). What is the relationship between (\ref{estvar-model-full}) and (\ref{estvar-model-partial})? \citet{ding-2021} provides a systematic treatment of this issue that relies on two ingredients. 
	
	First, that the vector of residuals from the multiple and partial regressions are exactly equal. Note that this was already proved by \citet{lovell-1963}, as we have seen above. This ingredient is important because the estimate of the covariance matrix of the regression error vector is a function of the regression residual vector. If the function depends \textit{only} on the vector of regression residuals, then we immediately have $\widehat{\Omega}_m=\widehat{\Omega}_p$. This already takes us some way in addressing the relationship between (\ref{estvar-model-full}) and (\ref{estvar-model-partial}).  
	
	The second ingredient relates to the other parts that make up the estimated covariance matrices in (\ref{estvar-model-full}) and (\ref{estvar-model-partial}). With reference to the multiple regression model in (\ref{model-full-sample})  and the partial regression model in (\ref{model-partial}), \citet[p.~6]{ding-2021} shows that  
	\begin{equation}\label{block-rel}
		\text{the } (2,1) \text{ block of } \left( W'W\right) ^{-1} W' = \left( {W^*_1}^{'} W^*_1\right)^{-1} {W^*_1}^{'},
	\end{equation}
	where $W = \left[ W_2 : W_1\right]$ and $W^*_1$ is defined (\ref{defw1star}).\footnote{This follows from a straightforward application of the inverse of partitioned matrices. Hence, I omit the proof.} 
	
	Bringing the two ingredients together, we get the following general result: (a) if the estimate of the error covariance matrix depends only on the vector of regression residuals, then (\ref{estvar-model-full}) and (\ref{estvar-model-partial}) are exactly equal; (b) if some degree of freedom correction is applied to generate the estimate of the error covariance matrix, then (\ref{estvar-model-full}) and (\ref{estvar-model-partial}) are equal up to the relevant degree of freedom correction; (c) if the estimate of the error covariance matrix depends, in addition to the residual vector, on the elements of the matrix of regressors, then (\ref{estvar-model-full}) and (\ref{estvar-model-partial}) are not, in general, equal even after making degree of freedom adjustments. 
	
	\subsection{The results}
	Using the above argument in different settings, \citet[Theorem~2, 3, 4]{ding-2021} shows the following: 
	\begin{enumerate}
		\item In models with homoskedastic errors, the covariance matrices from the multiple and partial regressions are equal up to a degree of freedom adjustment , i.e. $(N-k_1)\widehat{\Var} \left( \tilde{b}_1\right) = (N-k) \widehat{\Var} \left( b_1\right)$. This is because $\widehat{\Omega}_p = (1/N-k_1) \text{diag} \left[ \tilde{u}_i^2\right] $ and $\widehat{\Omega}_m = (1/N-k) \text{diag} \left[ u_i^2\right] $.
		
		\item In models with HC0 version of HCCM \citep[p.~299--300]{angrist-pischke-2009}, the covariance matrices from the multiple and partial regressions are exactly equal because $\widehat{\Omega}_p = \text{diag} \left[ \tilde{u}_i^2\right] $ and $\widehat{\Omega}_m = \text{diag} \left[ u_i^2\right] $.
		
		\item In models with HC1 version of HCCM \citep[p.~299--300]{angrist-pischke-2009}, the covariance matrices from the multiple and partial regressions are equal up to a degree of freedom adjustment because $\widehat{\Omega}_p = N/(N-k_1)\text{diag} \left[ \tilde{u}_i^2\right] $ and $\widehat{\Omega}_m = N/(N-k) \text{diag} \left[ u_i^2\right] $. Hence, $(N-k_1)\widehat{\Var} \left( \tilde{b}_1\right) = (N-k) \widehat{\Var} \left( b_1\right)$.
		
		\item In models using heteroskedasticity and autocorrelation (HAC) consistent covariance matrices \citep{newey-west-1987}, the covariance matrices from the multiple and partial regressions are exactly equal because $\widehat{\Omega}_p = \left( \omega_{|i-j|} \tilde{u}_i \tilde{u}'_j\right)_{1 \leq i, j, \leq N}  $ and $\widehat{\Omega}_m = \left( \omega_{|i-j|} \tilde{u}_i \tilde{u}'_j\right)_{1 \leq i, j, \leq N} $, as long as the same weights are used in both regressions.
		
		\item In models using the cluster robust estimate of the variance matrix (CRVE) \citep[p.~8]{cameron-miller-2015}, the covariance matrices from the multiple and partial regressions are exactly equal if no degree of freedom correction is used or if $G/(G-1)$ is used as the degree of freedom correction, where $G$ is the number of clusters. If $G(N-1)/(G-1)(N-k)$ is used as the degree of freedom correction, then $(N-k_1)\widehat{\Var} \left( \tilde{b}_1\right) = (N-k) \widehat{\Var} \left( b_1\right)$. 
	\end{enumerate}
	
	In all these cases, researchers can use estimated covariance matrices from the partial regression (\ref{model-partial}), with the relevant degree of freedom adjustment if necessary, to conduct proper statistical inference on the parameters from the multiple regression (\ref{model-full-sample}).

	\subsection{The cases that were left out}\label{sec:left-out}
	
	There are some cases where the estimate of the error covariance matrix depends, in addition to the residual vector, on elements of the covariate matrix. In these cases, (\ref{estvar-model-full}) and (\ref{estvar-model-partial}) will not be equal, even after making degrees of freedom corrections. Some common cases where this happens are: (a) models using HC2, HC3, HC4, or HC5 forms of HCCM; (b) some variants of the HAC and cluster-robust consistent covariance matrices. In these cases, it is not possible to use standard errors from the partial regression, with or without degrees of freedom correction, for inference about the coefficient vector in the multiple regression. But, it is still possible to compute the correct covariance matrix for the coefficient vector in the multiple regression without having to estimate that regression, as I now show.
	
	Let me discuss the HC2 case in detail; the other cases can be dealt with in a similar manner. Let $h_{ii}=W'_i\left(W' W\right)^{-1}W_i$, where $W_i$ is the $i$-th column of $W = \left[ W_2 : W_1\right] $ with reference to (\ref{model-full-sample}). The HC2 estimated covariance matrix of $b_1$ in (\ref{model-full-sample}) is given by  
	\begin{equation}\label{hc2-full}
		\text{the ($2,2$) block of }\left(W' W\right)^{-1} W' \frac{\text{diag}\left[ u_i^2\right]}{1-h_{ii}} W \left(W' W\right)^{-1}, 
	\end{equation}
	which, using (\ref{block-rel}), is 
	\begin{equation}
		\text{the ($2,2$) block of }
		\begin{pmatrix}
			* \\
			\left( {W^*_1}^{'} W^*_1\right)^{-1} {W^*_1}^{'}
		\end{pmatrix}
		\frac{\text{diag}\left[ u_i^2\right]}{1-h_{ii}} 
		\begin{pmatrix}
			{W^*_1} \left( {W^*_1}^{'} W^*_1\right)^{-1}  & *
		\end{pmatrix}, 
	\end{equation}
	and, therefore, is given by
	\begin{equation}
		\left( {W^*_1}^{'} W^*_1\right)^{-1} {W^*_1}^{'} \frac{\text{diag}\left[ u_i^2\right]}{1-h_{ii}} {W^*_1} \left( {W^*_1}^{'} W^*_1\right)^{-1},
	\end{equation}
	which, since $u_i^2$ can be replaced with $\tilde{u}_i^2$, is
	\begin{equation}\label{hc2}
	\left( {W^*_1}^{'} W^*_1\right)^{-1} {W^*_1}^{'} \frac{\text{diag}\left[ \tilde{u}_i^2\right]}{1-h_{ii}} {W^*_1} \left( {W^*_1}^{'} W^*_1\right)^{-1}.
	\end{equation}
	
	If we can estimate $h_{ii}$ using information available after estimating the partial regression (\ref{model-partial}), we can use (\ref{hc2}) to compute the covariance matrix that is necessary to conduct proper statistical inference on the coefficient vector $b_1$ in the multiple regression (\ref{model-full-sample}). In fact, using results on the inverse of partitioned matrices, we can do so. Hence, the following procedure for estimation and inference coefficient vector $b_1$ in the multiple regression (\ref{model-full-sample}) based on the estimation of the partial regression (\ref{model-partial}) can be suggested.
	
	\begin{enumerate}
		\item Compute $Y^* = \left[  I - W_2 (W'_2 W_2)^{-1} W'_2 \right]  Y$, and $W^*_1 = \left[  I - W_2 (W'_2 W_2)^{-1} W'_2 \right]  W_1$.
		
		\item Estimate the partial regression (\ref{model-partial}) by regressing $Y^*$ on $W^*_1$ and get the coefficient vector $\tilde{b}_1$.
		
		\item For $i = 1, 2, , \ldots, N$, compute $h_{ii}=W'_i\left(W' W\right)^{-1}W_i$ for the matrix of covariates, $W = \left[ W_2 : W_1\right]$, in the multiple regression (\ref{model-full-sample}). Let $W^{*}_2 = \left[  I - W_1 (W'_1 W_1)^{-1} W'_1 \right]  W_2$, and note that, using results for the inverse of partitioned matrices, we have
		\begin{equation}\label{inv-part}
		\left(W' W\right)^{-1} = \begin{pmatrix}
			W^{11} & W^{12} \\
			W^{21} & W^{22}
		\end{pmatrix},
		\end{equation} 
		where $W^{11} = \left( {W^*_2}^{'} W^*_2\right)^{-1}$, $W^{12}={W^{11}}^{'}= - \left( W'_2 W_2\right)^{-1}W'_2 W_1 \left( {W^*_1}^{'} W^*_1\right)^{-1}$, and $W^{22} = \left( {W^*_1}^{'} W^*_1\right)^{-1}$. Thus, we only compute $W^{11}$, $W^{12}$ and $W^{22}$, and then stack them to get $\left( W'W\right)^{-1} $. This avoids computing inverse of the $k \times k$ matrix $W'W$, and instead only involves computing inverses of $k_1 \times k_1$ or $k_2 \times k_2$ matrices.
		
		\item Use the matrix of covariates, $W^*_1$ and the vector of residuals, $\tilde{u}$, from the partial regression (\ref{model-partial}), to compute the covariance matrix of $\tilde{b}_1$ using (\ref{hc2}).
	\end{enumerate} 

	The main advantage of this procedure is computational. If a researcher were to estimate (\ref{model-full-sample}) and compute the covariance matrix of the full coefficient vector $b$, then she would have to program (\ref{hc2-full}), which would involve computing the inverse of a $k \times k$ matrix, $W'W$. If, instead, the partial regression-based procedure is used instead, then the researcher would: (a) compute $h_{ii}$ using (\ref{inv-part}), which involves inverses of $k_1 \times k_1$ and $k_2 \times k_2$ only; and (b) implement (\ref{hc2}), which involves computing the inverse of the $k_1 \times k_1$ matrix, ${W^*_1}^{'} W^*_1$. If $k_1$ and $k_2$ are much smaller than $k=k_1+k_2$, then there might be a significant reduction in computational burden. 
	
	The four step procedure outlined above can be used for HC3, HC4 variants of HCCM, and might even be applied to different variants of cluster-robust covariance matrices for which Theorem~3 and~4 in \citet{ding-2021} do not hold. The first step would remain unchanged; the second step would only change if any quantity other than, or in addition to, $h_{ii}$ is needed; only the third step would change significantly, where the researcher would need to use the correct expression for the relevant estimated covariance matrix, in place of (\ref{hc2}). 
	
	For instance, for HC3 and HC4 variants of heteroskedasticity consistent covariance matrices, steps~1, 2 and 3 would remain unchanged; in step~4, she would use either
	\begin{equation}\label{hc3}
		\left( {W^*_1}^{'} W^*_1\right)^{-1} {W^*_1}^{'} \frac{\text{diag}\left[ \tilde{u}_i^2\right]}{\left( 1-h_{ii}\right)^2 } {W^*_1} \left( {W^*_1}^{'} W^*_1\right)^{-1};
	\end{equation} 
	for HC3, or,
	\begin{equation}\label{hc4}
		\left( {W^*_1}^{'} W^*_1\right)^{-1} {W^*_1}^{'} \frac{\text{diag}\left[ \tilde{u}_i^2\right]}{\left( 1-h_{ii}\right)^{\delta_i} } {W^*_1} \left( {W^*_1}^{'} W^*_1\right)^{-1},
	\end{equation} 
	for HC4, where $\delta_i = \max \left( 4, N h_{ii}/k\right) $ and $k$ is the total number of regressors in (\ref{model-full-sample}), including the constant.
	
	\section{Discussion and conclusion}\label{sec:conclusion}
	
	By way of concluding this paper, I would like to offer two comments. My first comments is about the substance and usefulness of the YFWL theorem. In the context of a linear regression model, the YFWL theorem shows that if a researchers is interested in estimating only a subset of the parameters of the model, she need not estimate the full model. She can partition the set of regressors into two subsets, those that are of interest and those that are not of direct interest (but needs to be used for conditioning). She can regress the outcome variable on the conditioning set to create the residualized outcome variable. She can regress each of the variables of interest on the conditioning set to create corresponding residualized covariates. Finally, she can regress the residualized outcome variable on the residualized covariates of interest to get the estimated parameter of interest. 
	
	For statistical inference also, she can work with a partial regression in many, but not all, cases. If the errors are homoskedastic, then the estimated covariance matrix from the partial regression can be used for inference once a degree of freedom adjustment is made. If the errors are heteroskedastic, as is often the case in \textit{cross sectional} data sets, and the researcher wishes to use the HC0 or HC1 variant of HCCM, then she can use the estimated covariance matrix from the partial regression without any adjustment in the case of HC0 and with a degree of freedom adjustment in the case of HC1. If the researcher is using a \textit{time series} data set and wishes to use a HAC covariance matrix, then she can use the estimated covariance matrix from the partial regression as is. If the researcher is using a \textit{panel data} set and wishes to use the standard cluster robust covariance matrix, then she can use the estimated covariance matrix from the partial regression without any adjustment; if a degree of freedom correction is used to compute the cluster robust covariance matrix, then she will need to apply a corresponding degree of freedom adjustment. 
	
	If, on the other hand, the researcher wishes to use the HC2, HC3, HC4 or HC5 variants of heteroskedasticity consistent covariance matrix, or if she wishes to use other variants of the cluster robust covariance matrix or HAC covariance matrices, then she cannot use above results, i.e. the covariance matrices from multiple and partial regressions are no longer equal, even after degrees of freedom adjustments. Instead, the researcher can use the four-step computational method proposed in section~\ref{sec:left-out} if she wishes to use information from the partial regression to conduct proper statistical inference about parameters in the multiple regression.
	
	My second comment relates to a puzzle in the intellectual history of the YFWL theorem. The puzzle arises from noting that  \citet{udny-1907}'s result is essentially the same as the result proved by \citet{frisch-waugh-1933}, as I have demonstrated. What is puzzling, therefore, is that \citet{frisch-waugh-1933} do not cite \citet{udny-1907}. This omission seems puzzling to me given two fact. First, \citet{frisch-waugh-1933} refer to a previous paper by one of the authors: \citet{frisch-1929}. It is also to be noted that Frisch had published a related work in 1931 \citep{frisch-mudgett-1931}. What is interesting is that in both these papers, there is explicit reference to the partial correlation formula introduced by Yule. Recall that Yule had introduced a novel notation for representing partial correlations and coefficients in a multiple regression equation in his 1907 paper that I have discussed. The notation was then introduced to a wider readership in statistics with his 1911 book \citep{yule-1911}. This book was extremely popular and ran several editions, later ones with Kendall \citep{yule-kendall-1948}. Second, \citet{frisch-waugh-1933} use the exact same notation to represent the regression function that Yule had introduced in his 1907 paper. as I have highlighted. 
	
	These two facts suggest that Ragnar Frisch was familiar with Yule's work. In fact, Yule's work on partial correlations and multiple regression was the standard approach that was widely used and accepted by statisticians in the early part of the twentieth century \citep{griffin-1931}.  Therefore, it is a puzzle of intellectual history as to why \citet{frisch-waugh-1933} did not cite Yule's $ 1907 $ result about the equality of multiple and partial regression coefficients which was, in substantive terms, exactly what they went on to prove in their paper. Of course, \citet{frisch-waugh-1933} used a different method of proof from \citet{udny-1907}, as I have demonstrated in this paper. But in substantive terms, \citet{frisch-waugh-1933} proved the same result that Yule had established more than two decades ago. No matter what the reason for the omission in history, it seems that now is the right time to acknowledge Yule's pioneering contribution by attaching his name to a theorem that is so widely used in econometrics.
	
	\bibliographystyle{apalike}
	\bibliography{yfwl_refs}
	
	\begin{appendices}
		
		\section{Proof of Yule's result}\label{app:yule}
		
		The first step of the proof is to derive the normal equations arising from the least squares method of estimation. Given the random variables $x_1, x_2, \ldots, x_n$, which are demeaned versions of $X_1, X_2, \ldots, X_n$, the method of least squares chooses constants $b_1, b_2, \ldots, b_n$ to minimize
		\[
		\sum \left( x_1 - b_2 x_2 - b_3 x_3 - \cdots - b_n x_n\right)^2, 
		\]  
		where the sum runs over all observations in the sample. The first order conditions of this minimization problem are referred to as the `normal equations'. Differentiating the above with respect to $b_j$, we have
		\begin{equation}
			\sum x_j \left( x_1 - b_2 x_2 - b_3 x_3 - \cdots - b_n x_n\right) = 0 \quad \left( j=2, 3, \ldots, n\right), 
		\end{equation}
		so that, using Yule's notation, the normal equations are given by
		\begin{equation}\label{yule-norm-eq}
			\sum x_j x_{1.234 \ldots n} = 0 \quad \left( j=2, 3, \ldots, n\right), 
		\end{equation}
		which shows that the residual in the regression equation (\ref{lrg-1}) is uncorrelated with all the regressors included in the model, a result that holds for any regression function whose coefficients are estimated with the method of least squares.
		
		Now consider, $x_{2.34 \ldots n}$, the residuals from a regression of $x_2$ on $x_3, x_4, \ldots, x_n$, and note that it is a linear function of $x_2,x_3, x_4, \ldots, x_n$. Hence, using (\ref{yule-norm-eq}), we have
		\[
		\sum x_{1.234 \ldots n} x_{2.34 \ldots n} = \sum x_{1.234 \ldots n} \left( x_2 - c_1 x_3 - \cdots - c_{n-2} x_n \right)  = 0,
		\]
		for some constants $c_1, \ldots, c_{n-2}$. 
		
		We are now ready to prove the main result: coefficients from multiple and partial regressions are numerically the same.
		\begin{align*}
			0 = & \sum  x_{2.34 \ldots n} x_{1.234 \ldots n} \\
			= &  \sum x_{2.34 \ldots n} \left( x_1 - b_{12.34 \ldots n} x_2 + b_{13.24 \ldots n} x_3 + \cdots + b_{1n.234 \ldots n-1} x_n \right) \\
			= & \sum x_{2.34 \ldots n} \left( x_1 - b_{12.34 \ldots n} x_2\right) \\
			= & \sum x_{2.34 \ldots n} x_1 -  b_{12.34 \ldots n} \sum x_{2.34 \ldots n} x_2 \\
			= & \sum x_{2.34 \ldots n} \left( b_{13.4 \ldots n} x_3 + b_{14.3 \ldots n} x_4 + \cdots + b_{1n.34 \ldots n} x_n + x_{1.34\ldots n}\right) -  b_{12.34 \ldots n} \sum x_{2.34 \ldots n} x_2 \\
			= & \sum x_{2.34 \ldots n} x_{1.34\ldots n} -  b_{12.34 \ldots n} \sum x_{2.34 \ldots n} x_2 \\
			= & \sum x_{2.34 \ldots n} x_{1.34\ldots n} -  b_{12.34 \ldots n} \sum x_{2.34 \ldots n} \left( b_{23.4 \ldots n} x_3 + b_{24.3 \ldots n} x_4 + \cdots + b_{2n.34 \ldots n} x_n + x_{2.34\ldots n}\right) \\
			= & \sum x_{2.34 \ldots n} x_{1.34\ldots n} -  b_{12.34 \ldots n} \sum x_{2.34 \ldots n} x_{2.34\ldots n}
		\end{align*}
		Hence, we have (\ref{yule-1}):
		\[
		\frac{\sum \left( x_{1.34 \ldots n} \times x_{2.34 \ldots n}\right) }{\sum \left( x_{2.34 \ldots n}\right)^2 } = b_{12.34 \ldots n}.
		\]

		\section{Proof of Frisch and Waugh's result}\label{app:frisch}
		
		Consider (\ref{fw-2}) and let
		\begin{equation}\label{def-mij}
			m_{ij} = \sum x_i x_j \quad \left( i,j = 0, 1, 2, \ldots, n\right), 
		\end{equation}
		where the sum runs over all observations, and $x_0, x_1, \ldots, x_n$ are demeaned versions of $X_0, X_1, \ldots, X_n$.\footnote{\citet{chipman-1998} presents Frisch and Waugh's results using projection matrices. That is not correct. \citet{frisch-waugh-1933} did not use projection matrices in their proof, as I show in this section.} Then, \citet[p.~394]{frisch-waugh-1933} assert that the regression equation in (\ref{fw-2}) can be expressed as the following equation that specifies the determinant of the relevant ($n+1$)-dimensional matrix to be zero: 
		\begin{equation}\label{det-ne}
			\begin{vmatrix}
				x_0 & x_1 & \cdots & x_n \\ 
				m_{10} & m_{11} & \cdots & m_{1n} \\ 
				\vdots & \vdots & \ddots & \vdots \\ 
				m_{n0} & m_{n1} & \cdots & m_{nn}  
			\end{vmatrix} = 0.
		\end{equation}
		To see this, which \citet{frisch-waugh-1933} do not explain, perhaps because it was obvious to them, we need to recall two things. First, if we expand the determinant in (\ref{det-ne}) using the first row of the matrix, we will get an equation of the following form,
		\begin{equation}\label{dim:nplus1}
			a_0 x_0 + a_1 x_1 + \cdots + a_n x_n = 0,
		\end{equation}
		where $a_0$ is the determinant obtained by deleting the first row and first column (of the matrix whose determinant is being considered in (\ref{det-ne})), $a_1$ is $-1$ times the determinant obtained by deleting the first row and second column, $a_2$ is $1$ times the determinant obtained by deleting the first row and third column, and so on.\footnote{The signs alternate because the determinant obtained by deleting the first row and the $j$-th column is multiplied by $ -1^{(1+j)}$, where $j=1, 2, \ldots, n+1$. } Assuming $a_0 \neq 0$, which is guaranteed as long as the regressors are not perfectly collinear, this gives
		\begin{equation}\label{dim:nplus1-1}
			x_0  =  -\frac{a_1}{a_0} x_1 - \cdots  -\frac{a_n}{a_0} x_n.
		\end{equation}
		This has the same form as (\ref{fw-2}) and all we need to do is to show that the coefficients in (\ref{fw-2}) are what appears in (\ref{dim:nplus1-1}). To do so, we can use the normal equation and Cramer's rule. 
		
		Recall that the normal equations that had been written in (\ref{yule-norm-eq}) can, with reference to the least squares estimation (\ref{fw-2}), be written more compactly as $X'X b = X'y$, where $X = \left[ x_1 : x_2 : \cdots : x_n\right] $ is the matrix obtained by stacking the regressors column-wise, $y = x_0$ is the dependent variable, and $b$ is the least squares coefficient vector:
		\[
		b = \left[ b_{01.34 \ldots n} \quad b_{02.24 \ldots n} \quad \cdots \quad b_{0n.234 \ldots n-1}\right]. 
		\] 
		Note that the ($i,j$)-th element of $X'X$ is $m_{ij}$ as defined in (\ref{def-mij}), and the $i$-th element of $X'y$ is $m_{i0}$. Thus, the normal equations can be written as 
		\begin{equation}\label{sys-ne}
			\begin{bmatrix}
				m_{11} & m_{12} & \cdots & m_{1j} & \cdots & m_{1n} \\ 
				\vdots & \vdots & \cdots &  \vdots & \cdots  & \vdots \\ 
				m_{n1} & m_{n2} & \cdots & m_{nj} & \cdots & m_{nn}  
			\end{bmatrix} 
			\begin{bmatrix}
				b_{1} \\ 
				\vdots \\ 
				b_{n}  
			\end{bmatrix}
			=  \begin{bmatrix}
				m_{10} \\ 
				\vdots \\ 
				m_{n0}  
			\end{bmatrix}.
		\end{equation}
		Applying Cramer's rule \citep[p.~221]{strang-2006} to this equation system to solve the $b$ vector, keeping track of how many columns are switched and recalling that switching rows (columns) of a matrix only changes the sign of the determinant \citep[p.~203]{strang-2006}, we see that the coefficients in (\ref{fw-2}) and (\ref{dim:nplus1-1}) are identical. Thus, for $j=1,2, \ldots, n$
		\[
		b_j = \frac{\left| B_j \right| }{\left| A\right| },  
		\] 
		where
		\[
		A = \begin{bmatrix}
			m_{11} & m_{12} & \cdots & m_{1j} & \cdots & m_{1n} \\ 
			\vdots & \vdots & \cdots &  \vdots & \cdots  & \vdots \\ 
			m_{n1} & m_{n2} & \cdots & m_{nj} & \cdots & m_{nn}  
		\end{bmatrix} 
		\]
		and 
		\[
		B_j = \begin{bmatrix}
			m_{11} & m_{12} & \cdots & m_{10} & \cdots & m_{1n} \\ 
			\vdots & \vdots & \cdots &  \vdots & \cdots  & \vdots \\ 
			m_{n1} & m_{n2} & \cdots & m_{n0} & \cdots & m_{nn}  
		\end{bmatrix} 
		\]
		is obtained from $A$ by replacing the $j$-th column with $(m_{10} \quad \ldots \quad m_{n0})'$.

		Now consider (\ref{fw-3}) and let 
		\begin{equation}\label{def-mij}
			m'_{ij} = \sum x'_i x'_j \quad \left( i,j = 0, 1, 2, \ldots, n\right), 
		\end{equation}
		where the sum runs, once again, over all observations. Using the same logic as above, we will be able to see that the regression equation in (\ref{fw-3}) can be expressed as
		\begin{equation}\label{det-ne-1}
			\begin{vmatrix}
				x'_0 & x'_1 & \cdots & x'_{n-1} \\ 
				m'_{10} & m'_{11} & \cdots & m'_{1,n-1} \\ 
				\vdots & \vdots & \ddots & \vdots \\ 
				m'_{n-1,0} & m'_{n-1,1} & \cdots & m'_{n-1,n-1}  
			\end{vmatrix} = 0.
		\end{equation}
		
		The strategy will now be to use (\ref{det-ne}) and (\ref{det-ne-1}) to show that the first $n-1$ coefficients in (\ref{fw-2}) are equal to the corresponding coefficients in (\ref{fw-3}). The first thing is to relate $m_{ij}$ and $m'_{ij}$. This is easy to do by noting that 
		\begin{equation}
			x'_j = x_j - \frac{m_{jn}}{m_{nn}}x_n \quad \left( j = 0, 1, 2, \ldots, n-1\right),
		\end{equation} 
		because $x'_j$ is the residual from a regression of $x_j$ on $x_n$. Multiplying $x_i$ on both sides and summing over all observations for $x_i$ and $x_j$, we get
		\begin{equation}\label{mij}
			m'_{ij} = m_{ij} - \frac{m_{in} m_{jn}}{m_{nn}}.
		\end{equation}
		The second step is to return to (\ref{det-ne}) and carry out a series of elementary row operations that converts elements in the second through the penultimate row from $m_{ij}$ to $m'_{ij}$. From the row beginning with $m_{10}$ in (\ref{det-ne}) subtract $-m_{1,n}/m_{n,n}$ times the last row; from the row beginning with $m_{20}$ subtract $-m_{2,n}/m_{n,n}$ times the last row; and so on. Note that these row operations do not change the determinant of the matrix \citep[p.~204]{strang-2006}. Hence, these series of elementary row operations will convert (\ref{det-ne}) to 
		\begin{equation}\label{det-ne-2}
			\begin{vmatrix}
				x_0 & x_1 & \cdots & x_{n-1} & x_n \\ 
				m'_{10} & m'_{11} & \cdots & m'_{1,n-1} & 0 \\ 
				\vdots & \vdots & \ddots & \vdots & \vdots \\ 
				m'_{n-1,0} & m'_{n-1,1} & \cdots & m'_{n-1,n-1} & 0 \\ 
				m_{n0} & m_{n1} & \cdots & m_{n,n-1}& m_{nn}  
			\end{vmatrix} = 0.
		\end{equation}
		because of (\ref{mij}). Now consider the following determinant equation
		\begin{equation}\label{det-ne-3}
			\begin{vmatrix}
				x'_0 & x'_1 & \cdots & x'_{n-1} & 0 \quad \\ 
				m'_{10} & m'_{11} & \cdots & m'_{1,n-1} & 0 \quad \\ 
				\vdots & \vdots & \ddots & \vdots & \vdots \quad \\ 
				m'_{n-1,0} & m'_{n-1,1} & \cdots & m'_{n-1,n-1} & 0 \quad \\ 
				c_0 & c_1 & \cdots & c_{n-1}& 1  \quad
			\end{vmatrix} = 0,
		\end{equation}
		for some arbitrary constants $c_0, \ldots, c_{n-1}$, and by expanding the determinant by the last column, note that (\ref{det-ne-3})  is equivalent to (\ref{det-ne-1}). 
		
		Expanding (\ref{det-ne-2}) by the first row, we get
		\begin{equation}\label{eq:n}
			a_0 x_0 + a_1 x_1 + \cdots + a_{n-1} x_{n-1} + a_n x_n = 0
		\end{equation}
		and expanding (\ref{det-ne-3}) by the first row, we get
		\begin{equation}\label{eq:n-1}
			a'_0 x'_0 + a'_1 x'_1 + \cdots + a'_{n-1} x'_{n-1} = 0,
		\end{equation}
		where
		\begin{equation}\label{eq:coeff}
			a_i = a'_i m_{nn} \quad (i=0,1, 2, \ldots, n-1).
		\end{equation}
		Rearranging (\ref{eq:n}) as
		\begin{equation}
			x_0  =  -\frac{a_1}{a_0} x_1 - \cdots  -\frac{a_{n-1}}{a_0} x_{n-1} -\frac{a_n}{a_0} x_n,
		\end{equation}
		and rearranging (\ref{eq:n-1}) as
		\begin{equation}\label{dim:n-1}
			x'_0  =  -\frac{a'_1}{a'_0} x'_1 - \cdots  -\frac{a'_{n-1}}{a'_0} x'_{n-1},
		\end{equation}
		and noting that (\ref{det-ne-2}) is a re-expression of (\ref{fw-2}) while (\ref{det-ne-3}) is a re-expression of (\ref{fw-3}), we have
		\begin{equation}
			b_{0i.34 \ldots n} = b'_{0i.34 \ldots n} \quad (i=1, 2, \ldots, n-1).
		\end{equation}
		
		\section{Proof of Lovell's results}\label{app:lovell}
		
		Instead of following the proof in \citet{lovell-1963}, I will instead use \citet{greene} and \citet{ding-2021}. 
		
		\subsection{Parameter estimates are same}
		
		Let us start by recalling that the normal equations for (\ref{model-full-sample}), which we have seen before in in the context of Yule's proof and Frisch and Waugh's proof, can be written as $W'W b = W'Y$, where 
		\[
		W = \left[ W_2 : W_1\right] \text{ and } b' = \left[ b'_2 : b'_1\right]. 
		\]
		The normal equations can be expanded to give
		\begin{equation}
			\begin{pmatrix}
				W'_2 W_2 & W'_2 W_1 \\
				W'_1 W_2 & W'_1 W_1 
			\end{pmatrix}
			\begin{pmatrix}
				b_2 \\
				b_1
			\end{pmatrix} =
			\begin{pmatrix}
				W'_2 Y\\
				W'_1 Y
			\end{pmatrix},
		\end{equation}
		which, in turn, gives us the following two equation systems:
		\begin{align}
			W'_2 W_2 b_2 + W'_2 W_1 b_1 & = W'_2 Y \label{eq:first}\\
			W'_1 W_2 b_2 + W'_1 W_1 b_1 & = W'_1 Y. \label{eq:second}
		\end{align}
		From (\ref{eq:first}), we get 
		\[
		b_2 = \left( W'_2 W_2\right)^{-1} \left[ W'_2 Y - W'_2 W_1 b_1\right] 
		\]
		and substituting this in (\ref{eq:second}), and simplifying, we get
		\begin{equation}\label{bexpress-1}
			b_1 = \left( W'_1 M_{W_2} W_1\right)^{-1} W'_1  M_{W_2} Y,
		\end{equation}
		where
		\begin{equation}\label{def-P-M}
			M_{W_2} = I - P_{W_2}, \text{ and } P_{W_2} = W_2 \left( W'_2 W_2\right)^{-1} W'_2.
		\end{equation}
	
		Since $M_{W_2}$ is symmetric and idempotent, we can write (\ref{bexpress-1}) as
		\begin{equation}\label{bexpress-2}
			b_1 = \left( {W^*_1}^{'} W^*_1\right)^{-1} {W^*_1}^{'} Y^*,
		\end{equation}
		where $W^*_1 = M_{W_2} W_1$ is the matrix of column-wise stacked residuals from regressions of the columns of $W_1$ on $W_2$, and $Y^* = M_{W_2} Y$ is the vector of residuals from a regression of $Y$ on the $W_2$. Turning to (\ref{model-partial}), we see that
		\[
		\tilde{b}_1 = \left( {W^*_1}^{'} W^*_1\right)^{-1} {W^*_1}^{'} Y^*,
		\] 
		which establishes that $b_1 = \tilde{b}_1$.
		
		\subsection{Residuals are same}
		
		The residual vector from (\ref{model-full}) is given by $u = (I - P_W) Y$, and the residual vector from (\ref{model-partial}) is given by $\tilde{u} = (I - P_{W^*_1}) Y^* = (I - P_{W^*_1}) (I-P_{W_2}) Y$. Thus, if we can show that
		\begin{equation}\label{decomp}
			P_W = P_{W_2} + P_{W^*_1} + P_{W^*_1}P_{W_2}, 
		\end{equation}
		then the proof will be complete.
		
		We see that, with reference to (\ref{model-full}), 
		\begin{equation}
			P_W = W \left( W' W\right)^{-1} W' = 
			\begin{pmatrix}
				W_2 & W_1
			\end{pmatrix}
			\begin{pmatrix}
				W'_2 W_2 & W'_2 W_1 \\
				W'_1 W_2 & W'_1 W_1 
			\end{pmatrix}^{-1}
			\begin{pmatrix}
				W'_2 \\
				W'_1
			\end{pmatrix},
		\end{equation}
		which, using the inverse of partitioned matrices \citep[p.~993--994]{greene}, becomes
		\[
		P_W = \begin{pmatrix}
			W_2 & W_1
		\end{pmatrix}
		\begin{pmatrix}
			\left(  W'_2 W_2 - W'_2 W_1 \left( W'_1 W_1\right)^{-1} W'_1 W_2\right)^{-1}  & -\left( W'_2 W_2\right)^{-1} W'_2 W_1 F \\
			-F W'_1 W_2 \left( W'_2 W_2\right)^{-1} & F 
		\end{pmatrix}^{-1}
		\begin{pmatrix}
			W'_2 \\
			W'_1
		\end{pmatrix},
		\]
		where
		\[
		F = \left( W'_1 W_1 - W'_1 W_2 \left( W'_2 W_2\right)^{-1} W'_2 W_1\right)^{-1} = \left( W'_1 M_{W_2} W_1\right)^{-1}.
		\]
		Hence,
		\begin{align}\label{eq:simpl1}
			P_W & = W_2 \left( W'_2 W_2\right)^{-1} W'_2 + W_2 \left( W'_2 W_2\right)^{-1} W'_2 W_1 F W'_1 W_2 \left( W'_2 W_2\right)^{-1} W'_2 \nonumber \\
			& \quad - W_2 \left( W'_2 W_2\right)^{-1} W'_2 W_1 F W'_1 - W_1 F W'_1 W_2 \left( W'_2 W_2\right)^{-1} W'_2 + W_1 F W'_1,
		\end{align}
		Note that the last two terms in (\ref{eq:simpl1}) can be combined to give
		\[
		- W_1 F W'_1 W_2 \left( W'_2 W_2\right)^{-1} W'_2 + W_1 F W'_1 = W_1 F W'_1 \left[ I - P_{W_2}\right] = W_1 F W'_1 M_{W_2}, 
		\]
		and the second and third terms in (\ref{eq:simpl1}) can be combined to give
		\begin{align*}
			& W_2 \left( W'_2 W_2\right)^{-1} W'_2 W_1 F W'_1 W_2 \left( W'_2 W_2\right)^{-1} W'_2  \\
			& \qquad - W_2 \left( W'_2 W_2\right)^{-1} W'_2 W_1 F W'_1 = - W_2 \left( W'_2 W_2\right)^{-1} W'_2 W_1 F W'_1 M_{W_2}.
		\end{align*}
		Using these in (\ref{eq:simpl1}) we get
		\begin{align*}
			P_W & = W_2 \left( W'_2 W_2\right)^{-1} W'_2 + W_1 F W'_1 M_{W_2} - W_2 \left( W'_2 W_2\right)^{-1} W'_2 W_1 F W'_1 M_{W_2} \\
			& = P_{W_2} + \left[ I - W_2 \left( W'_2 W_2\right)^{-1} W'_2\right] W_1 F W'_1 M_{W_2} \\
			& = P_{W_2} + M_{W_2} W_1 F W'_1 M_{W_2} \\
			& = P_{W_2} + W^*_1 \left( {W^*_1}^{'} W^*_1\right)^{-1} {W^*_1}^{'},	
		\end{align*}
		because $F = \left( W'_1 M_{W_2} W_1\right)^{-1} = \left( W'_1 M'_{W_2} M_{W_2} W_1\right)^{-1} = \left( W^*_1 W^*_1\right)^{-1}$, where $W^*_1 = M_{W_2} W_1$, and $M_{W_2}$ is symmetric and idempotent. Hence, we have
		\[
		P = P_{W_2} + P_{W^*_1},
		\]
		and further that
		\[
		P_{W_2} P_{W^*_1} = W_2 \left( W'_2 W_2\right)^{-1} \underbrace{W'_2 W^*_1}_{=0} \left( {W^*_1}^{'} W^*_1\right)^{-1} {W^*_1}^{'} = 0,
		\]
		because
		\[
		W'_2 W^*_1 = W'_2 M_{W_2} W_1 = W'_2 \left( I - W_2 \left( W'_2 W_2\right)^{-1} W'_2 \right) W_1 = \left(W'_2-W'_2 \right) W_1 = 0 \times W_1 = 0.
		\]
		Since $P_{W_2}$ and $P_{W^*_1}$ are both symmetric, this implies that $P_{W^*_1}P_{W_2}=0$. Hence, we have
		\begin{equation}
			P_W = P_{W_2} + P_{W^*_1} + P_{W^*_1}P_{W_2},	
		\end{equation}
		and this establishes (\ref{decomp}).

	\end{appendices}
	
\end{document}